\documentclass{article}

\usepackage{PRIMEarxiv}

\usepackage[utf8]{inputenc} 
\usepackage[T1]{fontenc}    
\usepackage{hyperref}       
\usepackage{url}            
\usepackage{booktabs}       
\usepackage{amsfonts}       
\usepackage{nicefrac}       
\usepackage{microtype}      
\usepackage{lipsum}
\usepackage{fancyhdr}       
\usepackage{graphicx}       
\usepackage{xcolor}
\usepackage{todonotes}
\graphicspath{{media/}}     

\usepackage{listings}   
\definecolor{codegreen}{rgb}{0,0.6,0}
\definecolor{codegray}{rgb}{0.5,0.5,0.5}
\definecolor{codepurple}{rgb}{0.58,0,0.82}
\definecolor{backcolour}{rgb}{0.95,0.95,0.92}

\lstdefinestyle{mystyle}{
    backgroundcolor=\color{backcolour},   
    commentstyle=\color{codegreen},
    keywordstyle=\color{magenta},
    numberstyle=\tiny\color{codegray},
    stringstyle=\color{codepurple},
    basicstyle=\ttfamily\footnotesize,
    breakatwhitespace=false,         
    breaklines=true,                 
    captionpos=b,                    
    keepspaces=true,                 
    showspaces=false,                
    showstringspaces=false,
    showtabs=false,                  
    tabsize=2
}

\title{GrADyS-SIM - A OMNET++/INET simulation framework for Internet of Flying things}

\author{
  Thiago Lamenza, Marcelo Paulon, Breno Perricone, Bruno Olivieri,  Markus Endler  \\
  Laboratory for Advanced Collaboration, Departmento de Informática \\
  Pontificial Catholic University of Rio de Janeiro (PUC-Rio)\\
  Rio de Janeiro, Brazil\\
}

\begin{document}
\maketitle

\begin{abstract}
This technical report describes GrADyS-SIM, a  framework for simulating cooperating swarms of UAVs in joint mission in hypothetical landscape and communicating through RF radios. 
The framework was created to aid and verify the communication, coordination and context-awareness protocols being developed in the  GrADyS project. GrADyS-SIM uses the OMNeT++ simulation library and its INET model suite and and allows for addition  of modified/customized  versions of some simulated components, network configurations and vehicle coordination, so that new coordination protocols can be developed and tested through the framework.
The framework simulates UAV movement dictated by file containing some MAVLink instructions and
affected on the fly by different network situations. 
The UAV swarm’s coordination protocol emerges from individual
interactions between UAVs and has the objective of optimizing the collection of sensor data over an area. It also allows
for the simulation of some types of failures to test the protocol’s adaptability. Every node in the simulation is highly
configurable making testing different network topographies, coordination protocols, node hardware configurations and
more a quick task.
\end{abstract}


\section{Introduction}

When developing new network solutions, simulation is a powerful tool for testing, analyzing performance and evaluating scalability. This fact becomes more apparent when these networks have mobile air vehicles as some nodes. 
Simulation allows for multiple hypothesis checks, through easy setup of different network configurations and topologies and faster play of specific use case scenarios, all enabling much faster debugging and adjustments compared to a deployment and testing in the field. 
However, any simulation operates according to models (wireless transmisssion, mobility model, terrain model, etc.) and is therefore not completely precise or accurate. 
Therefore, we believe in the combined and interleaved use of simulation and field tests, and are particularly interested in identifying in which aspects simulations and field tests - of the same scenario - differ, and to what degree.

This work, more specifically the development of GrADyS-SIM is one of the tangible results of the research project entitled GrADyS \cite{gradys2021} (Ground-and-Air Dynamic sensors networkS). An important application of network technology is the creation of vehicle ad hoc networks, populated by mobile nodes that communicate with static nodes and other mobile nodes to implement some behavior. The GrADyS\footnote{http://www.lac.inf.puc-rio.br/index.php/gradys/} project was created to investigate the applications of these UAV swarm networks in the monitoring of remote, dangerous or hard-to-reach regions through the collection of sensor data using UAV swarms.
UAV or Aerial swarms are a specific type of multi-agent robotic systems with three-dimension freedom of movement and communicating through wireless communication.  In UAV swarms one of the biggest challenge is to ensure correct action/movement coordination of the UAVs, since this is usually done on the fly (literally) and must rely on intermittent and unreliable wireless connections\cite{Abdelkader:CRR:2021}.

This technical report describes the simulation framework created to aid and validate the project's development. Through the OMNeT++ simulation library and the INET model suite and with the creation and modification of simulated components, network configurations and vehicle coordination protocols are created and tested in this simulation framework.

The framework simulates UAV movement dictated by a simplified file containing some MAVLink instructions and affected on the fly by different network situations. The UAV swarm's coordination protocol emerges from individual interactions between UAVs and has the objective of optimizing the collection of sensor data over an area. It also allows for the simulation of some types of failures to test the protocol's adaptability. Every node in the simulation is highly configurable making testing different network topographies, coordination protocols, node hardware configurations and more a quick task. OMNeT++ uses a configuration file system that lets the developer easily create and keep track of these situations. A normal use case would be benchmarking the data collection performance of different coordination protocols with a different amount of UAVs on a waypoint mission.

\section{OMNET/INET++}
OMNeT++ is a discrete event simulator implemented as a component based C++ library. It primarily allows developers to build complex network simulations by creating and extending components that communicate with each other to implement behaviour. It is highly extensible and modular and the INET model suite takes advantage of that to provide a large component library to aid developers in implementing network simulations. These highly parameterized and extensible components provide solutions for simulating environments, a complete network stack, mobility and many other things. 

This environment provides the necessary tools for developing flexible simulations capable of representing different network situations with enough fidelity to serve as a tool for testing and validating coordination and network protocols and the network topographies needed to implement real-life mobile networks. Data collection and observation serve to rapidly test many variations in the network's implementation. The simulation can be easily configured to represent these variations by customizing component parameters and by the extension or creation of new components when needed.

\section{Architecture}

\begin{figure}
    \centering
    \includegraphics[width=1\textwidth, height=0.5\textwidth]{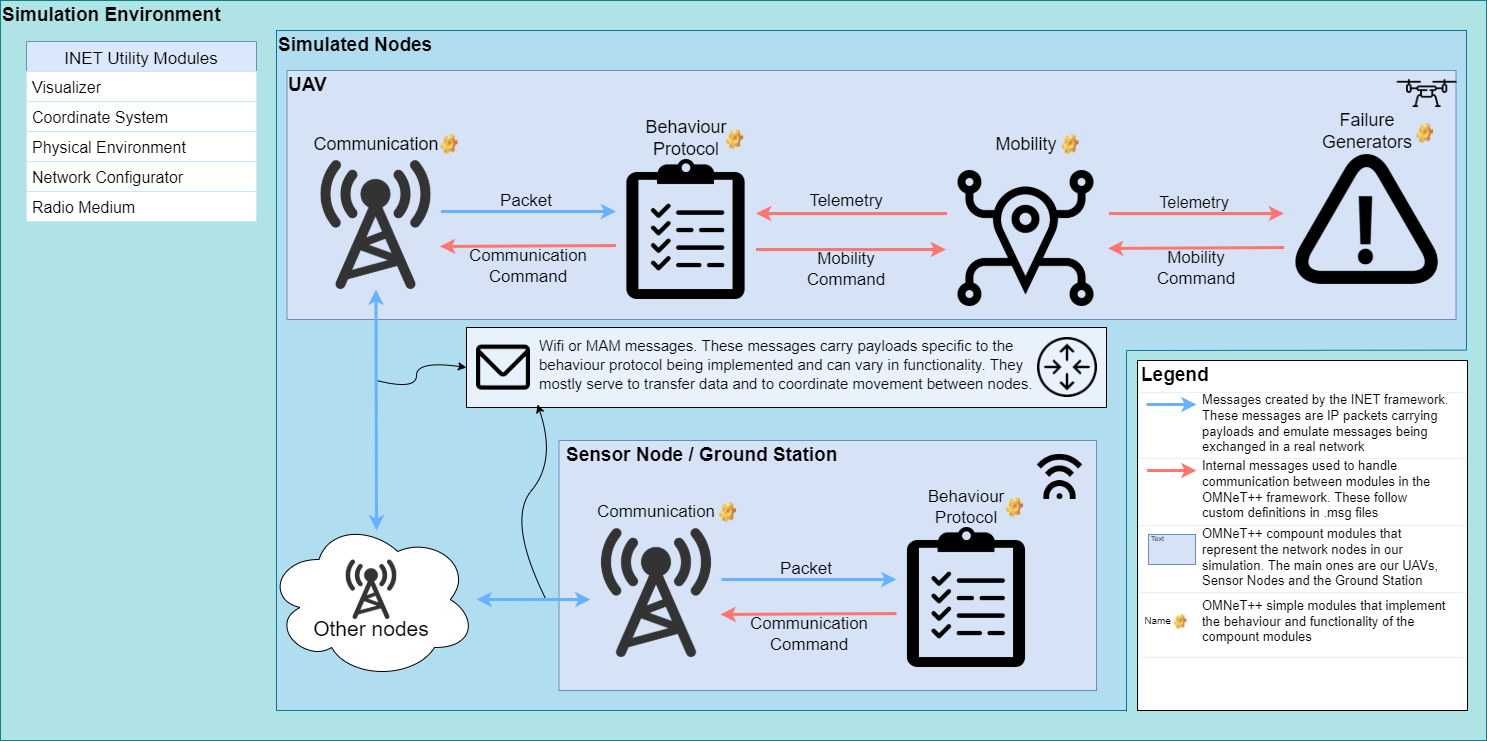}
    \caption{Project's Architecture Diagram}
\end{figure}

The framework's architecture is mainly composed of three models: one resposible for communication between UAVs (communication), one for controlling the node's movement (mobility) and the last to manage the interaction between the last two (protocol). The behaviour and implementation of these modules is detailed further below. They were made in such a way that the messages exchanged between them are sufficiently generic to allow the creation of a new protocol by creating a new protocol module, with no changes to the other ones by levaraging these generic messages to carry out different procedures. The messages exchanged between them are described on \verb|.msg| files like \verb|MobilityCommand.msg|, \verb|Telemetry.msg| and \verb|CommunicationCommand.msg| and define the format of the data being exchanged between them.

These three modules are loaded in a compound module defined by a \verb|.ned| file. In OMNeT++ \verb|.ned| files define modules that can use other modules forming a module tree. These modules can be simple (the leaves of the module tree) or a compound module that connects simple modules or other compound modules with gates. A network is a special kind of compound module that can be run as a simulation.

The compound module that represents UAVs in the simulation is the \verb|MobileNode.ned| and \verb|MobileSensorNode.ned| represents our sensors. These modules contain Communication and Mobility modules (defined in AdhocHost, this module's parent) and the Protocol module (defined in the file). The \verb|mobilityDrones.ned| file connects all the UAVs (called quads), sensors and some other modules necessary to the simulation.
\subsection{Components}
\subsubsection{Mobility}
The mobility module is responsible for controlling UAV movement and responding to requests from the protocol module to change that movement through MobilityCommand messages. It also needs to inform the protocol module about the current state of the UAV's movement through Telemetry messages. As part of the module initialization the waypoint list is attached to a Telemetry message so the protocol module has access to the tour the mobile node is following.

Currently the only mobility module being used is the \verb|DroneMobility.ned| module. The module responds to several commands defined the MobilityCommands message type that are used by the protocol module to control it. It's default behaviour is to follow a list of waypoints defined in a waypoint file but through commands it can perform several tasks like reversing course, going to a specific point or sitting idle. This module was adapted from the VehicleMobility module provided by INET to simulate ground vehicle movement through a series of waypoints. It was adapted to work in three dimensions, allowing flight, and to follow a waypoint file akin to the MAVLink waypoint file format. 

\begin{lstlisting}[language=C++, caption=MobilityCommand.msg]
// Commands that the mobility module should be capable of carrying out
enum MobilityCommandType {
    // Makes the UAV reverse on its course
    // No params
    REVERSE=0; 
    
    // Makes the UAV travel to a specific waypoint, following the tour pack
    // Param 1: Waypoint index
    GOTO_WAYPOINT=1;
    
    // Makes the UAV go to a specific coordinate and orient
    // itself so it can continue the tour afterwards
    // Param 1: x component of the coord
    // Param 2: y component of the coord
    // Param 3: z component of the coord
    // Param 4: Next waypoint(Waypoint the UAV should go to after reaching the target)
    // Param 5: Last waypoint(Waypoint the UAV used to reach the coords)
    GOTO_COORDS=2;
}

// Message declaration containing the command Id and its parameters 
message MobilityCommand {
    MobilityCommandType commandType;
    double param1=-1;
    double param2=-1;
    double param3=-1;
    double param4=-1;
    double param5=-1;
}
\end{lstlisting}

\begin{lstlisting}[language=C++, caption=Telemetry.msg]
// Activity that the UAV is currently carrying out
enum DroneActivity { 
    IDLE=0; 
    NAVIGATING=1;
    REACHED_EDGE=2; 
    FOLLOWING_COMMAND=3;
}

// Message declaration designed to share necessary UAV
// information with the communication module
message Telemetry {
    int nextWaypointID=-1;
    int lastWaypointID=-1;
    int currentCommand=-1;
    bool isReversed=false;
    DroneActivity droneActivity;
}
\end{lstlisting}

An optional feature of the mobility module is attaching a failure generator module. They connect to the mobility module using the same gates the protocol module does and use that to send commands in order to simulate failures. This can be used to trigger random shutdowns and even to simulate energy consumption. An example of a module that simulates energy consumption is the SimpleEnergyConsumption, a parametrized component to simulate consumption and battery capacity. It sends RETURN\_TO\_HOME messages to the vehicle when the UAV's battery reaches a certain threshold and shuts it down when the battery is depleted.
\subsubsection{Communication}
INET provides built in support for the simulation of real communications protocols and the communication module takes advantage of this to simulate communication between nodes. It also has to inform the protocol module of the messages being received by sharing the messages themselves and listen to orders from the protocol module through CommunicationCommands.

\begin{lstlisting}[language=C++, caption=CommunicationCommand.msg]
enum CommunicationCommandType {
   // Sets the payload that the communication module sends
   SET_PAYLOAD=0;
   // Sets the target of the communication message (null means broadcast)
   SET_TARGET=1;
}

// Message declaration for the communication command
message CommunicationCommand {
   CommunicationCommandType commandType;
   
   // Template for the SET_PAYLOAD message type 
   inet::FieldsChunk *payloadTemplate;
   
   // Target for the set target command
   string target;
}
\end{lstlisting}

The communication module has several used implementations. These implementations contain functions that interface with INET's communication capabilities but don't implement interaction with any other module. The interactions themselves are defined and controlled by the protocol module.

\subsubsection{Protocol}
The protocol module manages the interaction between the movement and communication of the mobile nodes. It makes use of the messages provided by it's two sibling modules to create node interaction strategies. It mostly reacts to messages it receives from those modules and determines which orders to give them to achieve the desired result.

It gathers information about the current state of the simulation by analysing Telemetry messages received from the Mobility module and Packets forwarded to it by the Communication module. An important task it performs is the definition of the message sent by the Communication module. These messages will be sent to other nodes that will themselves handle them. The messages are inserted into IP Packets as payload. They can have different formats depending on the protocol being implemented.

An example of a UAV coordination protocol that was implemented in this framework was DADCA\cite{bolivieri2020}. The DADCA investigates whether it is possible to implement a
distributed algorithm to coordinate several fully autonomous (i.e., non-human-controlled) UAVs collecting data from a WSN without centralized control or knowledge of internal UAV states and relying on only ad-hoc communication.

Another example of a protocol implemented was MAM\cite{paulon2021}. MAM proposes two alternatives to BTMesh's default relay algorithm ($MAM_0$ and $MAM_{\Delta}$) that may achieve higher packet delivery rates and lower energy draw when routing data towards a Mobile-Hub.

\begin{lstlisting}[language=C++, caption=Message exchanged in the Dadca protocol - DadcaMessage.msg]
enum DadcaMessageType
{
  HEARTBEAT = 0; 
  PAIR_REQUEST = 1; 
  PAIR_CONFIRM = 2;
  BEARER = 3;
}

class DadcaMessage extends FieldsChunk
{
  chunkLength = B(34); // Fixed chunk length
  int sourceID = -1;  // ID of the message's source
  int destinationID = -1; // ID of the message's destination
  int nextWaypointID = -1; // ID of the next waypoint
  int lastWaypointID = -1; // ID of the last waypoint
  int dataLength = 5; // Length of the imaginary data being carried in the message
  int leftNeighbours = 0; // Neighbours to the left of the UAV
  int rightNeighbours = 0; // Neighbours to the right of the UAV
  bool reversed = false; // Reverse flag which indicates the current direction the UAV is travelling in
  DadcaMessageType messageType = HEARTBEAT; // Type of message
}
\end{lstlisting}
Protocols implement the IProtocol interface and extend \verb|CommunicationProtocolBase.ned| which provides useful stub functions to use when implementing protocols.
\begin{lstlisting}[language=C++, caption=Function definitions for protocols - CommunicationProtocolBase.ned]
// Redirects message to the proper function
virtual void handleMessage(cMessage *msg);

// Handles package received from communication
// This packet is a message that was sent to the UAV
virtual void handlePacket(Packet *pk) {};

// Handles telemetry received from mobility
// The mobility module exchanges mobility information in the form of telemetry
virtual void handleTelemetry(Telemetry *telemetry) {};

// Sends command to mobility
virtual void sendCommand(MobilityCommand *order);
// Sends command to communication
virtual void sendCommand(CommunicationCommand *order);

// Sets a timeout
virtual void initiateTimeout(simtime_t duration);
// Checks if the module is timed out
virtual bool isTimedout();
\end{lstlisting}
Some of the currently implemented protocols are as follows:
\begin{itemize}
    \item \verb|ZigZagProtocol.ned| and \verb|ZigZagProtocolSensor.ned|
    
    These files implement the mobile node and the sensor side of the ZigZag protocol. This prococol manages a group of UAVs folowwing a set path passing above several sensors from where they pick up imaginary data from those sensors. The UAVs also interact with each other sending several messages to coordinate their movement.

    Heartbeat messages are sent on a multicast address, if these are picked up by sensors they respond with data. If they are picked up by other UAVs they initiate a communication pair by sending a Pair Request message which is them confirmed by the other UAV with a Pair Confirmation message. The UAV furthest away from the starting point of the path sends its data to the other UAV in the pair and they both reverse their movement. The objective is that over time the UAVs will each occupy an equally sized section of the course, picking up data on the way and sharing it at their section's extremities.
    
    \item \verb|DadcaProtocol.ned| and \verb|DadcaProtocolSensor.ned|
    
    This protocol is similar to the ZigZagProtocol. It also manages data collection by mobile nodes in a set path. The difference is that this method aims to speed up the process of equally spacing the UAVs in the course by implementing a more advanced movement protocol.

    When the Pair Confirmation message is recieved by both UAVs, confirming the pair, both UAVs take note of the number of neighours on their left (closer to the start) and their right (further from the start) and share this information with their pair. Both update their neighbour count and use it to calculate a point in the course that would represent the extremity of both their sections if their current count of neighbours is accurate. Them they both travel together to this point and revert. This behaviour is implemented with a sequence of commands that get queued on the mobility module.
\end{itemize}

\subsection{Operation}
The nature of the operation of a OMNeT++ simulation is dictated by the messages exchanged between modules and how they react to them. In the case of the GrADyS project the focus of our simulations is the use of quad copters to collect data from static sensors spread in a field. The UAVs and sensors are the network nodes of the simulation and are the main focus for development efforts. Each of these nodes are composed of three main modules, illustrated in the previous sections, that interact with each other to form the node's behaviour. 

Since all our nodes are composed of the same types of modules development is fast and simple and the implemented coordination protocols focus on creating complex behaviour emerging from individual actions taken by the nodes. In the protocols that we have developed every node of the same type is functionally identical and there is no coordinator but there still needs to be coordination to ensure efficient data collection. In the Dadca protocol this is achieved by the collection of information by each of the UAVs to gather a basic understanding of the layout and the distribution of other UAVs in the formation, namely by the counting and sharing of other neighbour UAVs they have encountered. The Zigzag protocol more primitively has the UAVs reverse every they encounter another one, with the vehicle farthest away from the ground station passing on the data to the one closest, ensuring it will eventually reach it.

The way that these nodes act is mainly determined by the implementation of their protocol module. It uses the other two modules as both sensors and actuators, gathering information about the network's state through information about the node's movement and messages received from other nodes and using this information to command the other two modules to perform the desired behaviour. An example of this is how in the previously described Zigzag coordination protocol the protocol module, on receiving communication from another UAV, compares the location visible in their message with it's own location by analysing telemetry received from the mobility module and uses all this information to decide if it should collect or send data to the other vehicle and commands the mobility module to invert the UAV's movement.

\begin{figure}
    \centering
    \includegraphics[width=1\textwidth, height=1\textwidth]{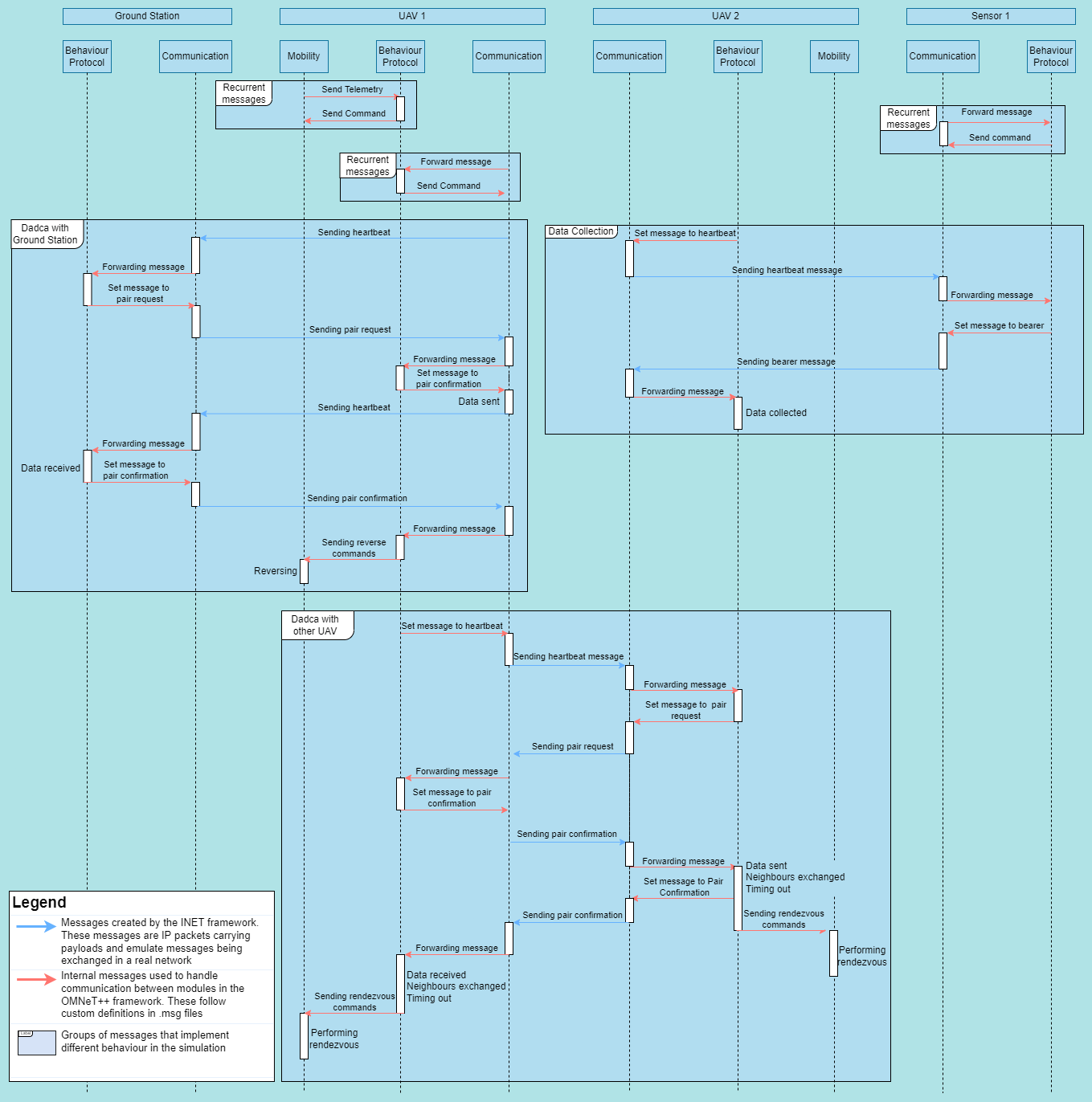}
    \caption{Event Diagram for the project's most common interactions}
\end{figure}

\section{Usage}
OMNeT++ simulations are initialized by .ini files. These files are used to set the many parameters of the simulation. These parameters control everything from the flight speed of the UAVs to the implementation of the protocol module to be used for the sensor nodes. They are neatly organized in these files and can be grouped in launch configurations. The launch configurations set a group of parameters for the simulation and can be used to easily switch between different sets of simulation types and node behaviours.

\begin{lstlisting}[language=C++, caption="Parameters used to set the number of UAVs and sensors in the simulation]
*.numUAVs = 2 // Initializes the *.quad[] array with 2 UAVs
*.numSensors = 8 // Initializes the *.sensors[] array with 8 sensors
\end{lstlisting}

\begin{lstlisting}[language=C++, caption=Parameters used to set and configure the quad's main modules]
// The protocol the UAV will follow (protocols explained further bellow)
// Change this to test other protocols like "ZigzagProtocol"
*.quads[*].protocol.typename = "DadcaProtocol" 

// The UAV's destination addresses (nodes it talks to and recieves messages from)
*.quads[0].app[0].destAddresses = "quads[1] sensors[0] 
                                   sensors[1] sensors[2] groundStation"

// Start time for the UAV's communication and mobility modules
// Change this to expertiment with different start timings
// The normal function gives a value between the two parameters
*.quads[1].app[*].startTime = normal(40s, 1s) 
*.quads[1].mobility.startTime = 40s

// The waypoint file the UAV should follow
*.quads[*].mobility.waypointFile = "paths/voo_ar.waypoints"
\end{lstlisting}

The mobilityDrones-omnetpp.ini file contains some launch configurations for Wifi only communication and shared Wifi and MAM communication, each with configurations for one to four UAVs. Launch configurations are defined in the same .ini file denoted by the [Config Sim2drone] tag where Sim2drone is the name of the launch configuration. The [Config Wifi] and [Config MAM] configs are base configs for the other ones and should not be ran.

\subsection{Installation}
In order to run the simulations and use the components in this repository you need to have both OMNeT++ and the INET framework installed.

Version 5.6 of OMNeT++ is required, to install it just follow these instructions. INET version 4.2 is also required, when first opening the OMNeT++ IDE you should be prompted with the option to install INET and all you need to do is accept it but if you need help check out the installation instructions.

After installing both OMNeT++ and INET you should be able to clone the repository to youw active OMNeT++ IDE workspace. To do this select File > Impor... then open the "git" section and select "Projects from git" then "Clone Uri". After that just fill in the URL for this repository and finish the process following the displayed instructions.

\subsection{Extension}
Since the project is Open Source anyone is able to download the source code, experiment and create new features as they see fit. The project's architecture was organized with extension in mind and it especially facilitates the development of new protocol modules, which are the main modules that should be implemented to develop new behavior for the network nodes.

After creating a new module all you need to do to test it is modifying the desired .ini configuration to load your protocol. The protocol module is flexible and can be loaded with any implemented protocol by changing it's typename. 

Developing a coordination protocol is one of the main tasks the simulation framework was created to solve. To showcase this the project includes instructions to develop a simple coordination module. This module simply directs each UAV to a waypoint and communicates with sensor nodes to collect data. No interaction between UAVs occurs, making this module not very useful but it still serves to showcase the project's capabilities as communication between all network nodes are implemented in the same way.

The first step is to create a new message. This message will be exchanged between network nodes to implement their behaviour. The message file contains the sender type (UAV, sensor or ground station) and the message's content, in this case a number representing a payload. 

Next ned definition files will be created for each of the network nodes. These definitions are accompanied by C++ definition files that script the node's reaction to each message. UAVs using this coordination module will react to messages received from sensors by collecting their payload. They will also constantly send out their current payload and wait for a response from a ground station, to which they will react by dropping their payload. Sensor nodes exclusively send out messages with a set payload and ground stations receive payloads from UAVs and acknowledge them.

The creation of all the ned and message files and C++ scripts is followed by the creation of a named configuration to facilitate the execution of this protocol. The named configuration will assign the coordination protocol to the simulation's nodes and will specify any necessary parameters. After the simulation is started the creted named configuration can be selected to run the developped scenario.

\section{Conclusion}

This work is a step within a set of deliverables for a project. The GrADyS project uses this tool to verify protocols and compare field test results with accurate sensors and UAVs.

This tool is in evolution, is licensed as open-source, and can be accessed freely\footnote{https://github.com/brunoolivieri/gradys-simulations}. From this, we hope that other research groups can reuse it with or without our involvement and contribute to that.

In our roadmap, we are already implementing a simulations` bind to the ArduPilot\footnote{https://ardupilot.org/} realistic vehicle simulator stack. This way, it will be possible to bring even more realism into the framework.

\section*{Acknowledgments}
This study was financed in part by AFOSR grant FA9550-20-1-0285.

\bibliographystyle{unsrt}  
\bibliography{references}

\end{document}